\documentclass[10pt, conference]{IEEEtran}
\IEEEoverridecommandlockouts
\usepackage[noadjust]{cite}
\usepackage{amsmath, amssymb, amsfonts, url}
\usepackage[ruled,vlined]{algorithm2e}
\usepackage{graphicx, tabularx, booktabs, subcaption}
\usepackage{textcomp}
\usepackage{xcolor}
\usepackage[acronym]{glossaries}

\def\BibTeX{{\rm B\kern-.05em{\sc i\kern-.025em b}\kern-.08em
    T\kern-.1667em\lower.7ex\hbox{E}\kern-.125emX}}

\begin{document}

\title{Robust Over-the-Air Computation with\\Type-Based Multiple Access
\thanks{This work is part of the project SOFIA PID2023-147305OB-C32 funded by MICIU/AEI/10.13039/501100011033 and FEDER/UE.}
}

\author{\IEEEauthorblockN{Marc Martinez-Gost\IEEEauthorrefmark{1}\IEEEauthorrefmark{2}, Ana Pérez-Neira\IEEEauthorrefmark{1}\IEEEauthorrefmark{2}\IEEEauthorrefmark{3}, 
Miguel Ángel Lagunas\IEEEauthorrefmark{2}}
\IEEEauthorblockA{
\IEEEauthorrefmark{1}Centre Tecnològic de Telecomunicacions de Catalunya, Spain\\
\IEEEauthorrefmark{2}Dept. of Signal Theory and Communications, Universitat Politècnica de Catalunya, Spain\\
\IEEEauthorrefmark{3}ICREA Acadèmia, Spain\\
\{mmartinez, aperez\}@cttc.es
}}

\newacronym{AI}{AI}{Artificial Intelligence}
\newacronym{AirComp}{AirComp}{over-the-air computation}
\newacronym{AWGN}{AWGN}{additive white Gaussian noise}
\newacronym{CNN}{CNN}{convolutional neural network}
\newacronym{CSI}{CSI}{channel state information}
\newacronym{DA}{DA}{direct aggregation}
\newacronym{DSB}{DSB}{double sideband}
\newacronym{FL}{FEEL}{federated learning}
\newacronym{FSK}{FSK}{frequency shift keying}
\newacronym{MSE}{MSE}{mean squared error}
\newacronym{NMSE}{NMSE}{normalized mean squared error}
\newacronym{PPM}{PPM}{pulse position modulation}
\newacronym{TBMA}{TBMA}{type-based multiple access}
\newacronym{SNR}{SNR}{signal-to-noise ratio}

\maketitle
\begin{abstract}
This paper utilizes the properties of type-based multiple access (TBMA) to investigate its effectiveness as a robust approach for over-the-air computation (AirComp) in the presence of Byzantine attacks, this is, adversarial strategies where malicious nodes intentionally distort their transmissions to corrupt the aggregated result. Unlike classical direct aggregation (DA) AirComp, which aggregates data in the amplitude of the signals and are highly vulnerable to attacks, TBMA distributes data over multiple radio resources, enabling the receiver to construct a histogram representation of the transmitted data. This structure allows the integration of classical robust estimators and supports the computation of diverse functions beyond the arithmetic mean, which is not feasible with DA. Through extensive simulations, we demonstrate that robust TBMA significantly outperforms DA, maintaining high accuracy even under adversarial conditions, and showcases its applicability in federated learning (FEEL) scenarios. Additionally, TBMA reduces channel state information (CSI) requirements, lowers energy consumption, and enhances resiliency by leveraging the diversity of the transmitted data. These results establish TBMA as a scalable and robust solution for AirComp, paving the way for secure and efficient aggregation in next-generation networks.
\end{abstract}

\section{Introduction}
The advent of highly complex and heterogeneous networks has created an urgent need for high-speed data transfer protocols capable of handling massive data volumes while enabling real-time processing and data-driven decision-making. To address these demands, future communication networks must integrate distributed computation directly into the communication process, as seen in applications like \gls{AI}, which require continuous aggregation of vast data from distributed devices to train and update models. These massive data flows call for efficient and scalable techniques, such as \gls{AirComp}, which leverages the waveform superposition property of wireless channels to aggregate data directly over the air. By aligning simultaneous transmissions to compute a desired function of the transmitted signals, \gls{AirComp} minimizes latency and radio resource usage, making it particularly relevant for large-scale distributed systems like wireless sensor networks and \gls{FL}, where high-speed data aggregation is crucial.

While \gls{AirComp} offers significant advantages in terms of efficiency and scalability, it also introduces critical security concerns \cite{perez2024waveforms}. Since the receiver does not access individual device transmissions, \gls{AirComp} inherently preserves privacy, which is a desirable feature in many applications. However, this same characteristic makes it challenging to identify malicious behavior, as the aggregated result conceals individual contributions. 
Consequently, \gls{AirComp} is particularly vulnerable to Byzantine attacks, where adversaries deliberately manipulate their transmissions to corrupt the aggregated outcome. Addressing these vulnerabilities is essential to ensure the robustness and reliability of \gls{AirComp} in adversarial environments.

The existing literature on Byzantine attacks and outlier detection in distributed computing has largely focused on general frameworks \cite{vempaty2013byzantine}, with limited attention to their application in \gls{AirComp}. The only proposed solution under the \gls{AirComp} framework is presented in \cite{sifaou2022robustfl}, where the authors introduce an algorithm operating between the physical and link layers. Specifically, devices are divided into subsets, which transmit data concurrently within each group. Then, the geometric median across groups is computed as a robust estimate of the mean. However, this approach is limited to \gls{FL} scenarios and only considers the computation of the arithmetic mean. Additionally, the algorithm is not fully in the physical layer, as it requires scheduling transmissions among users. Another line of research explores robust precoders for \gls{AirComp} \cite{hu2024multirelay, zhang2024channeluncertain, zhou2024otfs, an2021ris}, addressing channel imperfections by accounting for channel uncertainty in a probabilistic sense. However, these works do not address resiliency to adversarial attacks, leaving a critical gap in the robustness of \gls{AirComp}.

All previous works on \gls{AirComp} operate under a \gls{DA} framework, where the additive property of the wireless channel is used to achieve superposition, resulting in the receiver obtaining a single aggregated signal. To achieve resiliency against attacks with \gls{DA} in the physical layer, the function estimate has to be robust. However, the performance is heavily dependent on the data distribution and the nature of the attacks, limiting their effectiveness.  

In this work, we propose leveraging a \gls{TBMA} for \gls{AirComp} instead. In \gls{TBMA}, devices transmit in a single radio resource (i.e., time, frequency or code) according to their data, allowing the receiver to construct a histogram representation of the transmitted data. It has been previously shown the benefits of \gls{TBMA} over \gls{DA} in terms of power consumption and \gls{CSI} \cite{perez2024waveforms}. In this work we show that the diversity provided by multiple radio resources can be exploited at the receiver to identify and isolate attacks targeting specific resources. Moreover, \gls{TBMA} supports a broader range of function computations compared to \gls{DA}, as it decouples the attack detection and compensation from the function computation.

The contributions of this work are as follows:
\begin{enumerate}
    \item We propose a novel physical-layer framework for robust aggregation in \gls{TBMA}-based \gls{AirComp}, enhancing resilience against adversarial attacks.
    \item \gls{TBMA} enables the integration of both parametric and non-parametric estimators, providing flexibility and generality. 
    \item Unlike most works that assume the sample average, our scheme allows the computation of robust estimates for a variety of functions, which we evaluate.
    \item  We compare \gls{TBMA} with \gls{DA} and demonstrate the effectiveness of robust aggregation with \gls{TBMA} in a \gls{FL} use case, highlighting its practical applicability.
\end{enumerate}

The remaining part of the paper proceeds as follows:
Section II introduces aggregation techniques for \gls{AirComp} and presents the system model. Section III introduces robust techniques for \gls{TBMA} and section IV assesses their performance in different scenarios. Section V concludes the paper.

\section{System Model}
Consider a wireless network with $K$ distributed devices (i.e., transmitters) and a single server (i.e., receiver). Each device $k\in\mathcal{K}$ has local data $s_k$ (e.g., sensor measurements or local computations) and the goal of the network is to compute a function of the data $f(s_1,\dots,s_K)$. For this purpose, \gls{AirComp} is exploited to efficiently use the communication resources.

The traditional approach for \gls{AirComp} is encoding the information in the amplitude of the transmitted waveform. Assuming perfect synchronization and channel compensation, the additive nature of the wireless channel aggregates the signals at the input of the receiver. To compute the sample average, the receiver only needs to divide the amplitude of the received waveform by the number of transmitters. To compute other functions, the transmitters and receiver need to process the transmitted and received signals, respectively, a procedure termed nomographic function representation \cite{goldenbaum13_nomographic}.
These techniques, either analog or digital, fall under the umbrella of \gls{DA}, as the aggregation occurs over a single communication resource.
Alternatively, for robust \gls{AirComp} we propose using \gls{TBMA}.

\subsection{Type-based multiple access}
In \gls{TBMA} \cite{mergen2006tbma} there is a one-to-one mapping between $L$ different measurements and $L$ orthogonal
radio resources. For the sake of simplicity we assume that $s_k\in [1,\dots,L]$, which can be achieved via linear or nonlinear mappings (e.g., quantization).
Thus, in \gls{TBMA}, radio resources are allocated according to data, not users, enabling superposition when users have the same measurement.
The signal transmitted by user $k$ is $\varphi(s_k)$. For instance, in \eqref{eq:examples_tbma}, \gls{TBMA} can be implemented with $M$-ary \gls{FSK} when there is a bijective mapping between data and frequency resources; alternatively, \gls{TBMA} can be implemented with \gls{PPM}, where there is a bijective mapping between the data and the time shifts of a pulse.
\begin{equation}
\varphi(s_k) =
\begin{cases}
a_k e^{j2\pi s_kn/T} &
\text{for $M$-\gls{FSK}}\\
a_k e^{j2\pi (n-s_k)/T} &
\text{for \gls{PPM}}
\end{cases}
\label{eq:examples_tbma}
\end{equation}

In \eqref{eq:examples_tbma}, $T$ is the duration of the signal and $n=0,\dots,{N-1}$ is the discrete time index. The amplitude $a_k\in\mathcal{C}$ is used to compensate channel imperfections. Although these are two examples of how \gls{TBMA} can be implemented, this technique always requires an orthogonal support of signals. Furthermore, notice that each user uses a single from the $L$ available radio resource. 

At the receiver side, the signal at resource $\ell\in[1,\dots,L]$ is
\begin{equation}
y_\ell = 
\sum_{k\in\mathcal{K}} h_{k\ell}
\varphi(s_k) + w_\ell,
\end{equation}
where $h_{k\ell}\in\mathcal{C}$ is the channel between user $k$ and the receiver at resource $\ell$, and $w_\ell$ is the \gls{AWGN} sample at resource $\ell$ with power $\sigma^2$. Notice that $\varphi(s_k)\neq 0$ only when $s_k=\ell$.
We assume perfect \gls{CSI}, which can be achieved via channel inversion as $a_k=h_k^*/|h_k|^2$. We refer the reader to \cite{mergen2006tbma} for the consideration of channel knowledge and effects on error estimation, as the purpose of this paper is to show how \gls{TBMA} provides robustness towards Byzantine attacks. Additional techniques for channel compensation can be designed on top of the proposed design.

Computing the matched filter for the set of $y_{\ell}$ and normalizing by $K$ results in
\begin{equation}
    \textbf{r} = \frac{1}{K} [K_1,\dots,K_{L}]^T + [\tilde{w}_1,\dots,\tilde{w}_{L}]^T = \textbf{p} + \tilde{\textbf{w}}, 
    \label{eq:empirical_measure}
\end{equation}
where the $\ell$-th entry in $\textbf{p}$ corresponds to the fraction of devices $K_\ell/K$ that transmitted at resource $\ell$ (i.e., $s_k=\ell$) and $\tilde{w}_\ell$ is Gaussian noise with power $\tilde{\sigma}^2 = \sigma^2/K^2$. Provided that \mbox{$K>L$}, which is reasonable in \gls{AirComp} scenarios,  $\textbf{r}$ corresponds to a nosy version of the empirical measure $\textbf{p}$, this is, a histogram or type of the distributed data $s_k$.

From the noisy type $\mathbf{r}$, many of the functions of interest can be computed by aggregating the $L$ signals with an aggregation function $\Psi$. For instance,
\begin{equation}
\Psi(\mathbf{r}) =
\begin{cases}
\frac{1}{K} \sum_{\ell=1}^L 
\ell r_\ell &
\text{for arithmetic mean}\\
\exp\left(\frac{1}{K} \sum_{\ell=1}^L r_\ell \ln
\ell\right)
&
\text{for geometric mean}
\end{cases}
\label{eq:examples_aggregation}
\end{equation}

As mentioned previously, amplitude-based \gls{AirComp} requires non-linear pre-processing and post-processing to compute functions beyond the arithmetic mean. Conversely, \gls{TBMA} computes the function over the type $\mathbf{r}$. Besides \eqref{eq:examples_aggregation}, other functions such as the minimum or maximum reduce to classical detection problems for \gls{TBMA}.
Furthermore, from the communication perspective, one of the most relevant benefits of \gls{TBMA} over \gls{DA} is the saving in energy consumption. For instance, to compute the geometric mean with \gls{DA}, each device transmits $\ln(s_k)$, which may have a large impact in the transmitted power. Conversely, as shown in \eqref{eq:examples_aggregation}, \gls{TBMA} only chooses the radio resource and the transmitted power is independent of the function. 

The figure of merit is the \gls{NMSE} between the distributed data and the function estimated at the receiver side:
\begin{equation}
\text{NMSE} = \frac{|f(s_1,\dots,s_K)-\Psi(r_1,\dots,r_L)|^2}{f(s_1,\dots,s_K)^2}
\label{eq:figure_merit_mse}
\end{equation}

\subsection{Byzantine attacks}
In a Byzantine attack there is a set of users that try to corrupt the aggregation by sending malicious data. In the context of \gls{AirComp} and, particularly, in \gls{TBMA} we define a set of users $\mathcal{M}$ that attack a specific radio resource to alter the distribution of $\mathbf{p}$. We assume that the attackers are coordinated to attack the same radio resource $\tilde{\ell}$, which represents the worst case scenario in terms of \eqref{eq:figure_merit_mse}. We define the received signal in the presence of attackers as
\begin{equation}
\tilde{y}_\ell = 
\sum_{k\in\mathcal{K}} h_{k\ell}
\varphi(s_k) +
\sum_{m\in\mathcal{M}} h_{m{\ell}}
\varphi({\ell})
+ w_\ell,
\label{eq:attacked_type}
\end{equation}
and the corrupted type as
\begin{align}
\tilde{\textbf{r}} = &\frac{1}{K} [K_0,\dots,K_{L-1}]^T + 
\frac{1}{K} [0,\dots,M_{\tilde{\ell}},\dots,0]^T +\tilde{\mathbf{w}},
\label{eq:empirical_measure_corrupted}
\end{align}
where $M_{\tilde{\ell}}$ is the number of attackers at resource $\tilde{\ell}$. Without loss of generality we assume that $M\leq K$, as allowing the number of attackers to exceed the number of legitimate devices would result in most of the data being controlled by attackers, leading to a completely compromised system.
Although not considered in this work, attackers can be more severe if they adjust the transmission power (e.g., $|a_m|^2=P_{max}$, this is, transmitting at maximum power).

\section{Robust techniques for \gls{TBMA}}
The additional consumption of wireless resources of \gls{TBMA} with respect to \gls{DA} ―in fact, by a factor of $L$― provides more information at the receiver side. According to \eqref{eq:empirical_measure_corrupted}, when the number of attackers is small, we have $\tilde{\textbf{r}}\approx\textbf{r}$ and the \gls{NMSE} is mostly affected by channel noise. Conversely, when the number of attackers is large, radio resource $\tilde{\ell}$ can be easily spotted due to the structure imposed by \gls{TBMA}.
We will use this information to provide detection and compensation techniques against Byzantine attacks. Notice that this robustness is not only limited to attacks, but also to any abnormal alternation in the amplitude of the received type, such as channel fading. We leave the extension beyond attacks for future work.

Assume, without loss of generality, $f$ to be the arithmetic mean. Given $\mathbf{p}$ with an arbitrary distribution, attackers need to select an adequate resource $\tilde{\ell}$ to displace the mean. This implies that $\tilde{\ell}\leq \min{s_k}$ or $\tilde{\ell}\geq \max{s_k}$ are suitable candidates. However, in \gls{TBMA}, these attacks are very easy to detect over $\tilde{\mathbf{r}}$ because they represent outliers in the data distribution.
Thus, robust estimation over \gls{TBMA} reduces to the classical signal processing problem of outlier detection and robust estimation over data distributions. Many off-the-shelf techniques can be used to provide robustness. In the following we list a few:
\begin{itemize}
    \item \textit{Robust estimators}: for instance, the median is a non-parametric robust estimator of the arithmetic mean.
    \item \textit{Percentile truncation}: keeping only the data in a certain percentile range removes extreme outliers.
    \item \textit{Resampling techniques}: methods like bootstrapping or RANSAC \cite{fischler1981ransac} resample the data several times to robustly estimate the underlying distribution.
\end{itemize}

The purpose of this paper is not to provide a superior robust algorithm for \gls{TBMA}, but to show that classical techniques can be implemented over \gls{TBMA}. We define $\hat{\mathbf{r}}$ as the corrected type and we present an approach for robust estimation over \gls{TBMA} that integrates several techniques in Algorithm \ref{alg:algo}. First, we threshold the type below the noise level (e.g., $\theta_1=3\sigma^2$); then, percentile truncation is designed to remove extreme outliers that lie far outside the main distribution of the data. We define $P_1$ and $P_2$ as the low and high percentiles of $\tilde{\mathbf{r}}$, respectively.
The last step targets outliers that are not far from the overall data range but deviate significantly from their local context (i.e., their immediate neighbors). Under the assumption that a single resource is attacked, we identify values that are much larger than both their preceding and succeeding neighbors. Instead of removing these points, we compensate for their effect by replacing them with the average of their adjacent values. This ensures smoother transitions in the data while preserving the overall structure of the distribution. These operations can be executed in parallel across the $L$ radio resources, ensuring an efficient scaling with the number of resources.

From the corrected type $\hat{\mathbf{r}}$, the desired function can be estimated using the aggregation function $f\approx \hat{\Psi(\mathbf{r})}$. However, note that some information from non-attackers transmitting on resource $\tilde{\ell}$ may be lost from the non-attackers that were transmitting at the attacked resource. Unlike \gls{DA}, where the design of robust estimation is inherently coupled with the specific function being computed, \gls{TBMA} decouples attack detection and compensation from function computation. For instance, with \gls{DA}, a robust estimator for the arithmetic mean might be the median, requiring pre-processing and post-processing steps that are tightly linked to the mean computation. In contrast, \gls{TBMA} makes the outlier removal independent from the function estimation. This decoupling enables \gls{TBMA} to support the computation of multiple functions from a single transmission, offering greater flexibility compared to \gls{DA} methods.

It is critical to emphasize that the design of this algorithm and the corresponding parameters must be tailored to the specific distribution of $\mathbf{p}$. 
Additionally, the algorithm must account for the nature of potential attacks on the aggregation process. For example, if the attack strategy injects subtle but coordinated outliers within the main distribution, step 3 (local outlier compensation) becomes vital. Conversely, if the attack introduces extreme values far from the distribution, step 2 (percentile truncation) is more critical. By adapting the algorithm to both the data distribution and the anticipated attack strategies, it can robustly ensure the integrity of the aggregated results.

The authors in \cite{sifaou2022robustfl} tackle the problem of robust aggregation in amplitude-based \gls{AirComp}, which aligns with the general framework of \gls{TBMA}. In their approach, devices are grouped into clusters to disperse the attackers, with each group transmitting concurrently using \gls{DA} at different time steps. The resulting aggregations are robustly averaged using the geometric median. Notice that assigning different transmission times to groups is simply an orthogonal radio allocation, similar to what is done in \gls{TBMA}. The key distinction lies in how radio resources are assigned: in \cite{sifaou2022robustfl}, this is done randomly, whereas in \gls{TBMA}, it is based on the data. Moreover, the structure set by \gls{TBMA} allows compensating a greater number of attackers.

\begin{algorithm}[t]
\caption{Robust estimation over \gls{TBMA}}
\SetAlgoLined
\KwIn{$\tilde{\mathbf{r}}$, $\theta_1$, $\theta_2$, $P_1$ and $P_2$}
\KwOut{$\Psi(\hat{\mathbf{r}})$}
\vspace*{4 pt}
1. Threshold the noise:
\begin{equation}
\hat{r}_{\ell} =
\begin{cases}
\tilde{r}_{\ell} &
\text{if $|\tilde{r}_{\ell}|\geq\theta_1$}\\
0 &
\text{if $|\tilde{r}_{\ell}|<\theta_1$}
\end{cases}\nonumber
\end{equation}

2. Percentile truncation:
\begin{equation}
\hat{r}_{\ell} =
\begin{cases}
\tilde{r}_{\ell} &
\text{if $P_1\leq \tilde{r}_{\ell}\leq P_2$}\\
0 &
\text{otherwise}
\end{cases}\nonumber
\end{equation}

3. Outlier detection:
\begin{equation}
\hat{r}_{\ell} =
\begin{cases}
\tilde{r}_{\ell} &
\text{if $|\tilde{r}_{\ell}-\tilde{r}_{\ell-1}| \leq \theta_2$} \\
& \text{and $|\tilde{r}_{\ell}-\tilde{r}_{\ell+1}| \leq \theta_2$}, \\
\frac{\tilde{r}_{\ell-1}+\tilde{r}_{\ell+1}}{2} &
\text{otherwise.}
\end{cases}
\nonumber
\end{equation}

4. Function estimation: 
$f\approx \Psi(\hat{\mathbf{r}})$

\label{alg:algo}
\end{algorithm}

\section{Results}

\subsection{Comparison of TBMA with DA}

\begin{figure}[t]
    \centering
    \begin{subfigure}[b]{\columnwidth}
        \centering
        \includegraphics[width=\textwidth]{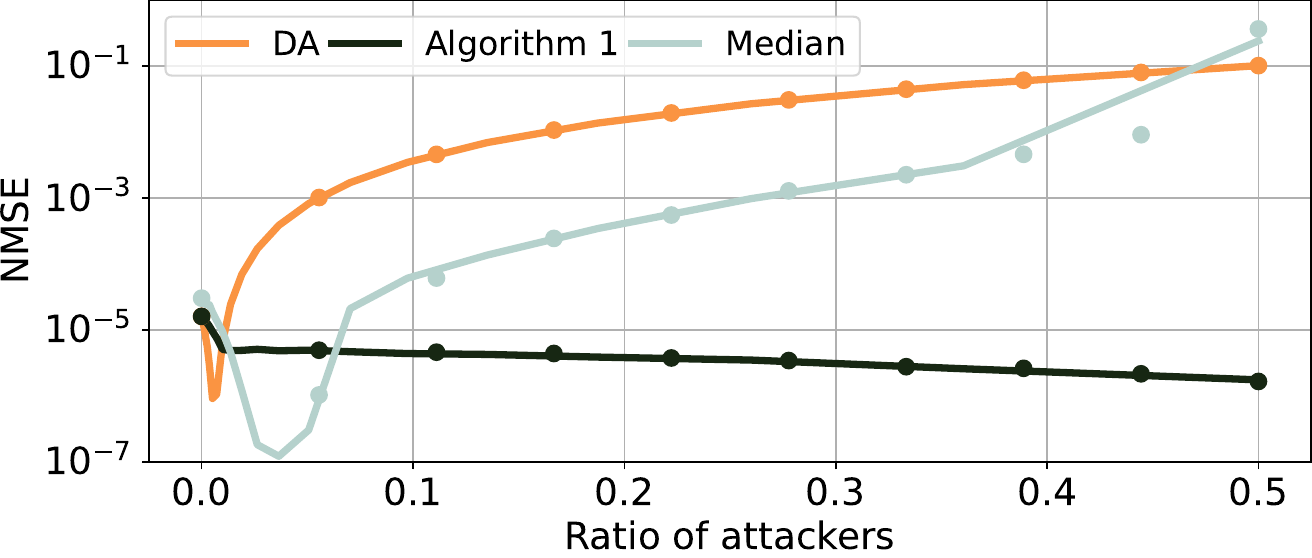}
        \caption{Arithmetic mean.}
        \label{fig:arith_gauss}
    \end{subfigure}
    
    \begin{subfigure}[b]{\columnwidth}
    \vspace{0.3cm}
        \centering
        \includegraphics[width=\textwidth]{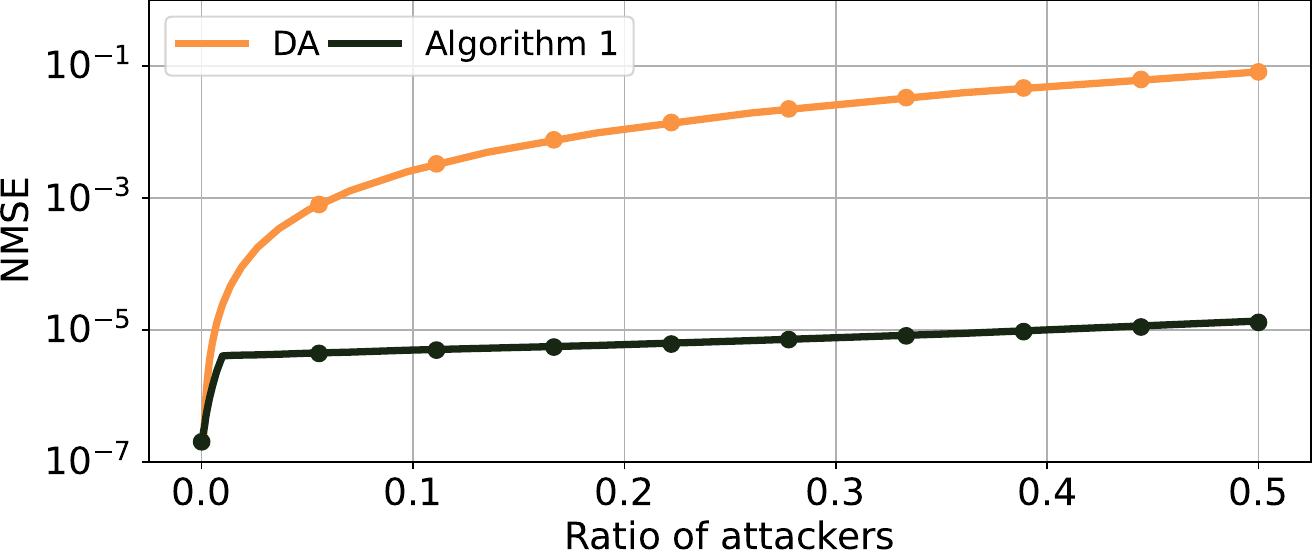}
        \caption{Geometric mean.}
        \label{fig:geo_gauss}
    \end{subfigure}
    \caption{\Gls{NMSE} for different \gls{AirComp} techniques at $\text{SNR}=30$ dB (lines) and 5 dB (markers).}
    \label{fig:tbma_vs_da}
\end{figure}

In Figure \ref{fig:tbma_vs_da} we evaluate different \gls{AirComp} techniques for different functions and data distributions under the influence of Byzantine attacks.
We focus on the arithmetic mean, which serves as the classical function in \gls{AirComp} and is widely used in aggregation tasks. Additionally, we compute the geometric mean as an example of an alternative function that is not present in the literature.
To ensure adequate density for \gls{TBMA} to represent data distributions effectively, the setup includes $K=10^4$ devices and $L=256$. In the succeeding section we will showcase \gls{TBMA} in a low density scenario. Different ratios of attackers are tested, and results are averaged over $10^3$ independent experiments.

Three methods are compared in the evaluation. The first is \gls{DA}, which aggregates all received signals without applying any robust estimation, making it highly vulnerable to attacker influence. We do not include any robust variations of \gls{DA} because no physical layer technique exists that can inherently provide robustness, even for the arithmetic mean.
The second approach is the robust \gls{TBMA} of Algorithm \ref{alg:algo}, for which we set $\theta_1=3\sigma^2, \theta_2=5, P_1=0.01 \text{ and } P_2=0.99$.
The third method computes the median from the \gls{TBMA} type, which serves as a robust estimate of the arithmetic mean. However, this approach is specific to the arithmetic mean and cannot be easily generalized to other functions. Notably, computing the median with \gls{DA} is not straightforward. 
The underlying modulation for \gls{DA} is \gls{DSB} modulation, while for \gls{TBMA} is \gls{PPM}, and we consider an \gls{AWGN} channel.

As shown in Figure \ref{fig:tbma_vs_da}, the results are consistent across different \gls{SNR} regimes, as the large number of signal aggregations provides robustness against channel noise.
Regarding the attacks, \gls{DA} is not robust, and the error increases with the number of attackers for all functions.
\gls{TBMA} with the median function is more robust than \gls{DA}, but the \gls{NMSE} still increases with the number of attackers. This emphasizes the need for robust \gls{AirComp} techniques, not just robust function estimates.
In this respect, Algorithm \ref{alg:algo} is resilient to attacks, and the error does not increase significantly with the ratio of attackers. For the arithmetic mean, the \gls{NMSE} even decreases because it is easier to detect outliers in the type. This effect is not observed in the geometric mean, as it involves nonlinear transformations.

\subsection{\gls{FL} use case}
\Gls{FL} is a distributed approach where multiple devices collaboratively train a global model by sharing model updates (e.g., gradients in a neural network), instead of raw data. \gls{AirComp} complements \gls{FL} by using the superposition property of the wireless channel to perform aggregation directly during transmission. This eliminates the need for separate aggregation at the receiver, reducing communication overhead and latency.

We consider the deployment of a \gls{FL} system in which a server coordinates the learning process of $K=50$ devices. The task is image classification using the MNIST dataset \cite{mnist}, consisting in images of handwritten digits ranging from 0 to 9. The learning model consists in a classical \gls{CNN} (see \cite{gost2023fskfl}) and deployed with the Flower framework \cite{flower}. Each device trains a local \gls{CNN} with a portion of the data and sends the parameters after each epoch. For transmission, each parameter is quantized, modulated with M-\gls{FSK} and transmitted in a \gls{TBMA} fashion. The receiver receives a type for each parameter and computes the arithmetic mean. Notice that an orthogonal set of resources is required for every parameter.
We consider $6\%$ of the devices ($M=3$) to generate Byzantine attacks and the figure of merit is the accuracy, this is, the number of correctly classified samples.

\begin{figure}[t]
    \centering
    \includegraphics[width=1\columnwidth]{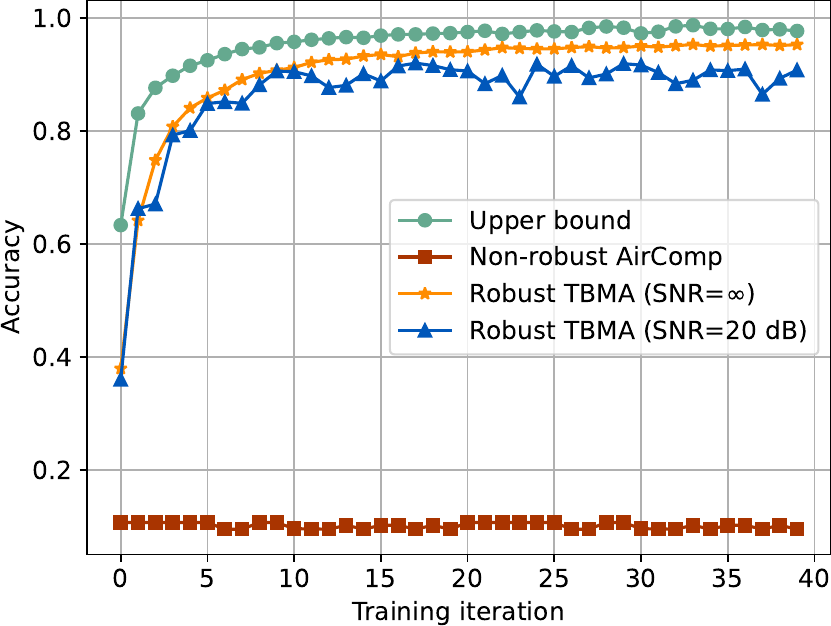}
  \caption{Test accuracy across training for an \gls{AirComp}-based \gls{FL} system under Byzantine attacks.}
  \label{fig:feel}
\end{figure}

Figure \ref{fig:feel} shows the convergence of the \gls{AirComp}-assisted \gls{FL} system. The upper bound represents the system trained in a noiseless channel and without attacks. In practice, even a small number of attackers is enough to corrupt the system and prevent convergence. \gls{DA} (i.e., non-robust \gls{AirComp}) achieves only $10\%$ accuracy, equivalent to random guessing given 10 classes. In contrast, robust \gls{TBMA} performs almost as good as the non-attacked system in the noiseless scenario. The performance is slightly below the baseline due to information loss from devices sharing resources with attackers. At high SNR, robust \gls{TBMA} maintains strong performance, as M-\gls{FSK} proves more resilient than \gls{PPM}, thanks to the frequency-based nature of the modulation.

\section{Conclusion}
This paper presents the first physical-layer robust aggregation method for \gls{AirComp} under Byzantine attacks. Unlike classical \gls{DA} techniques, which aggregate data in the amplitude domain, we propose using \gls{TBMA}, where orthogonal radio resources are allocated to different data, not transmitters. With \gls{TBMA}, the receiver obtains a histogram of the transmitted data, enabling straightforward detection of attacks. \gls{TBMA} allows the integration of classical robust estimators and outlier detection methods, extending beyond the sample average to support robust estimation of diverse functions that cannot be computed with \gls{DA}. Through simulations, we demonstrated that robust \gls{TBMA} significantly outperforms \gls{DA} in the presence of Byzantine attacks, maintaining high performance even with a substantial number of attackers. Additionally, we showcased its practical applicability in a \gls{FL} use case, highlighting its resilience and adaptability. These results establish \gls{TBMA} as a robust and general solution for \gls{AirComp} in adversarial settings. Future work will explore extending robust \gls{TBMA} to address other amplitude-related challenges, such as channel alterations.

\bibliographystyle{IEEEbib}
\bibliography{refs}

\begin{thebibliography}{10}

\bibitem{perez2024waveforms}
Ana P{\'e}rez-Neira, Marc Martinez-Gost, Alphan {\c{S}}ahin, Saeed Razavikia, Carlo Fischione, and Kaibin Huang,
\newblock ``Waveforms for computing over the air,''
\newblock {\em arXiv preprint arXiv:2405.17007}, 2024.

\bibitem{vempaty2013byzantine}
Aditya Vempaty, Lang Tong, and Pramod~K. Varshney,
\newblock ``Distributed inference with byzantine data: State-of-the-art review on data falsification attacks,''
\newblock {\em IEEE Signal Processing Magazine}, vol. 30, no. 5, pp. 65--75, 2013.

\bibitem{sifaou2022robustfl}
Houssem Sifaou and Geoffrey~Ye Li,
\newblock ``Robust federated learning via over-the-air computation,''
\newblock in {\em 2022 IEEE 32nd International Workshop on Machine Learning for Signal Processing (MLSP)}. IEEE, 2022, pp. 1--6.

\bibitem{hu2024multirelay}
Changjie Hu, Quanzhong Li, Qi~Zhang, and Qiang Li,
\newblock ``Robust design for multi-carrier multi-relay assisted over-the-air computation networks with bounded channel uncertainties,''
\newblock {\em IEEE Wireless Communications Letters}, vol. 13, no. 8, pp. 2290--2294, 2024.

\bibitem{zhang2024channeluncertain}
Hongrui Zhang, Xiao Tang, Ruonan Zhang, Turlykozhayeva Dana, Nurzhan Ussipov, and Zhu Han,
\newblock ``Distributionally robust over-the-air computation in presence of channel uncertainties,''
\newblock in {\em 2024 IEEE Wireless Communications and Networking Conference (WCNC)}, 2024, pp. 1--6.

\bibitem{zhou2024otfs}
Dongkai Zhou, Jing Guo, Siqiang Wang, Zhong Zheng, Zesong Fei, Weijie Yuan, and Xinyi Wang,
\newblock ``Otfs-based robust mmse precoding design in over-the-air computation,''
\newblock {\em IEEE Transactions on Vehicular Technology}, 2024.

\bibitem{an2021ris}
Qiaochu An, Yong Zhou, and Yuanming Shi,
\newblock ``Robust design for reconfigurable intelligent surface assisted over-the-air computation,''
\newblock in {\em 2021 IEEE Wireless Communications and Networking Conference (WCNC)}, 2021, pp. 1--6.

\bibitem{goldenbaum13_nomographic}
Mario Goldenbaum, Holger Boche, and Sławomir Stanczak,
\newblock ``Harnessing interference for analog function computation in wireless sensor networks,''
\newblock {\em IEEE Transactions on Signal Processing}, vol. 61, no. 20, pp. 4893--4906, 2013.

\bibitem{mergen2006tbma}
G.~Mergen and L.~Tong,
\newblock ``Type based estimation over multiaccess channels,''
\newblock {\em IEEE Transactions on Signal Processing}, vol. 54, no. 2, pp. 613--626, 2006.

\bibitem{fischler1981ransac}
Martin~A Fischler and Robert~C Bolles,
\newblock ``Random sample consensus: a paradigm for model fitting with applications to image analysis and automated cartography,''
\newblock {\em Communications of the ACM}, vol. 24, no. 6, pp. 381--395, 1981.

\bibitem{mnist}
Li~Deng,
\newblock ``The {MNIST} database of handwritten digit images for machine learning research,''
\newblock {\em IEEE Signal Processing Magazine}, vol. 29, no. 6, pp. 141--142, 2012.

\bibitem{gost2023fskfl}
Marc Martinez-Gost, Ana Pérez-Neira, and Miguel~Ángel Lagunas,
\newblock ``Frequency modulation aggregation for federated learning,''
\newblock in {\em GLOBECOM 2023 - 2023 IEEE Global Communications Conference}, 2023, pp. 1878--1883.

\bibitem{flower}
Daniel~J. Beutel, Taner Topal, Akhil Mathur, Xinchi Qiu, Javier Fernandez-Marques, Yan Gao, Lorenzo Sani, Kwing~Hei Li, Titouan Parcollet, Pedro Porto~Buarque de~Gusm{\~a}o, and Nicholas~D. Lane,
\newblock ``Flower: A friendly federated learning framework,''
\newblock Open-Source, mobile-friendly Federated Learning framework, Mar. 2022.

\end{thebibliography}

\end{document}